\documentclass[journal]{IEEEtran}
\usepackage{cite}
\usepackage{amsmath,amssymb,amsfonts}
\usepackage{textcomp}
\usepackage{xcolor}
\usepackage[hidelinks]{hyperref}
\usepackage{cleveref}
\usepackage{amsthm}
\newtheorem{theorem}{Theorem}
\newtheorem{proposition}{Proposition}

\newtheorem{lemma}{Lemma}
\theoremstyle{definition}
\newtheorem{definition}{Definition}
\theoremstyle{remark}
\newtheorem{remark}{Remark}
\DeclareMathOperator{\Tr}{Tr}
\def\BibTeX{{\rm B\kern-.05em{\sc i\kern-.025em b}\kern-.08em
T\kern-.1667em\lower.7ex\hbox{E}\kern-.125emX}}
\begin{document}

\title{Sufficient Algebras for Conditional Double Markovity}
\author{Masahito Hayashi and Jinpei Zhao%
\thanks{M. Hayashi is with the School of Data Science, The Chinese University
of Hong Kong, Shenzhen, China; the International Quantum Academy, Shenzhen,
China; and the Graduate School of Mathematics, Nagoya University, Japan
(e-mail: hmasahito@cuhk.edu.cn).}%
\thanks{J. Zhao is with the Department of Information Engineering, The
Chinese University of Hong Kong, Shatin, Hong Kong, China
(e-mail: jinpei@ie.cuhk.edu.hk).}%
\thanks{M. Hayashi was supported in part by the Guangdong Provincial Quantum
Science Strategic Initiative (Grant No. GDZX2505003) and the General R\&D
Projects of 1+1+1 CUHK-CUHK(SZ)-GDST Joint Collaboration Fund (Grant No.
GRDP2025-022).}
\thanks{J. Zhao was supported by the Hong Kong Research Grants Council (RGC) under grant numbers 14213826, and 14309924, and by the 1+1+1 CUHK–CUHK(SZ)–GDST Joint Collaboration Fund (Grant No.
GRDP2025-022).}}
\maketitle

\begin{abstract}
For four finite-valued classical variables, the double Markov
conditions $I(A;C\mid BD)=I(A;B\mid CD)=0$ give a label $J$
determined by each of $BD$ and $CD$ with $I(A;BC\mid DJ)=0$.
We study the same two conditions for a finite-dimensional quantum
state $\rho_{ABCD}$. Measurement outcomes on $A$ give a family
of conditional states on $BCD$ and marginal families on $BD$ and
$CD$. The minimal sufficient algebra of a family retains the
information needed for one channel to recover every state in
that family. The two regional minimal sufficient algebras have
the same image on the support of $\rho_{BCD}$ under the respective
partial-trace adjoints. This common algebra is the minimal
sufficient algebra of the joint family.

An isometry on the support of $\rho_{BCD}$ leaves $A$ unchanged
and gives an output state $\Omega$ on $ABCDJ$. For fixed $J$ and
a fixed isometry with output classical on $J$, the condition
$I(A;BC\mid JD)_\Omega=0$ is equivalent to an injective
representation of the common algebra on the support of
$\Omega_{JD}$ that is block diagonal on $J$ and intertwines the
isometry. With the quantum output spaces fixed and $J$ and the
isometry allowed to vary, such an output exists exactly when
the factor on which the algebra acts in each Koashi--Imoto
block has dimension at most that of the output system $D$.

Under the double Markov conditions, if the state is diagonal
in product bases on $B,C,D$, either $BD$ or $CD$ can record a
common label while preserving the original $ABCD$ marginal. For one-dimensional $D$, separate projective
measurements on $B$ and $C$ produce the same recorded state with
the original $ABC$ marginal and zero conditional mutual
information given $J$. A four-qubit example has one bit of
conditional mutual information $I(A;BC\mid DJ)$ in every extension
that preserves the original $ABCD$ marginal, and an isometry
achieves zero conditional mutual information on its output.
\end{abstract}


\section{Introduction}\label{sec:introduction}
\subsection{Double Markovity and common information}
For finite-valued classical random variables, the double Markov
conditions $I(A;C\mid BD)=I(A;B\mid CD)=0$ imply that there is a
common label $J=g_{BD}(B,D)=g_{CD}(C,D)$ almost surely with
$I(A;BC\mid DJ)=0$. One may choose $J$ to be the conditional law of
$A$ given $BD$. For constant $D$, this is the classical double-Markov
lemma \cite[Lemma~1]{HayashiZhao2026DoubleMarkov}.

The full information shared by $BD$ and $CD$ is described by
\[
 \mathcal G=\overline{\sigma(B,D)}\cap\overline{\sigma(C,D)},
\]
where the bars denote completion under the joint probability law.
The double Markov conditions are equivalent to conditional
independence of $A$ and $(B,C)$ given $\mathcal G$
\cite[Theorem~7.1]{Dawid1980ConditionalOperations}
\cite[Proposition~2.2]{SanMartinMouchartRolin2005CommonInformation}.
For constant $D$, this shared information can be recorded by a
label $K$ obtained from a bipartite graph. Its two vertex sets
consist of the positive-probability values of $B$ and of $C$.
Join $b$ to $c$ when $P(B=b,C=c)>0$ and assign one label to each
connected component, consisting of vertices linked by paths.
Either observed value determines the component of the pair.
Every common label must take the same value at the endpoints
of each edge, and hence along each path. Thus $K$ records all
the shared information, as in G\'acs--K\"orner common information
\cite{GacsKorner1973CommonInformation}. The conditional-law label
$J$ retains the shared information needed to recover the
conditional law of $A$. Its algebra can be strictly smaller
than the full shared algebra.

Now let a quantum state $\rho_{ABCD}$ satisfy the same double
Markov conditions. The two observations are $BD$ and $CD$, with
$D$ their shared subsystem. The two zero conditional mutual
informations give recovery of $\rho_{ABCD}$ from
$\rho_{ABD}$ by a channel on $BD$ and from $\rho_{ACD}$ by a channel
on $CD$, respectively \cite{Petz1986SufficientSubalgebras,HJP,Shirokov2016}.
Can one record a classical $J$ while preserving the original $ABCD$
marginal and obtain $I(A;BC\mid DJ)=0$? The example in
Section~\ref{sec:four-cycle-main} shows that this can fail.

\subsection{Sufficient algebras}
A measurement outcome of positive probability on $A$ defines a
conditional state on $BCD$. Taking its marginals on $BD$ and $CD$
gives two families indexed by the same outcomes. An observation is
sufficient if one channel recovers every joint conditional state
from the corresponding observed marginal. The minimal sufficient
algebra of a family is the smallest observable algebra from
which one channel can recover the entire family.

Theorem~\ref{thm:main-common-information} identifies the minimal
sufficient algebras of the two marginal families with that of
the joint family on $BCD$. To compare these algebras, a $BD$
observable acts on $BCD$ by the identity on $C$, and projection
on both sides gives its action on the support of $\rho_{BCD}$.
Section~\ref{subsec:common-sufficient-algebra} specifies the maps
for $BD$ and $CD$. For each observable in the joint family's
minimal sufficient algebra, each marginal algebra contains an
observable with that same supported image. We call the joint
algebra the common sufficient algebra. The theorem also gives a
conditional expectation onto this algebra whose trace-adjoint
channel preserves $\rho_{ABCD}$ when applied to $BCD$. The proof
uses the sufficiency theory of Jen\v{c}ov\'a and Petz
\cite[Theorems~1--2 and~5]{JencovaPetz2006Sufficiency}.

The block form of the common algebra comes from the Koashi--Imoto
decomposition \cite{KoashiImoto2002}. In the construction of
Section~\ref{subsec:isometric-reduction}, $J$ records the block
and $F$ holds the factor on which the algebra acts within that
block. Theorem~\ref{thm:classical-quantum-realization} characterizes
when observables in the common algebra can be represented by
operators on the output $JD$ that intertwine an isometry on
the support of $\rho_{BCD}$. For a fixed
isometry with output $\Omega$ classical on $J$, this is equivalent
to $I(A;BC\mid JD)_\Omega=0$. With $J$ and the isometry allowed to
vary, each factor carrying the algebra in a block must fit in
the output system $D$, and this condition also suffices.

The four-qubit example in Section~\ref{sec:four-cycle-main} gives
$I(A;BC\mid DJ)_\Omega=1$ for every $\Omega_{ABCDJ}$ whose $ABCD$
marginal is $\rho_{ABCD}$. Its common algebra is $M_2$, and the
observables on physical $D$ span a two-dimensional subspace of
$M_2$ after compression to the support of $\rho_{BCD}$. An
isometry to output systems $B',C',J,D'$ with a qubit $D'$ and
one-dimensional $J$ gives $I(A;B'C'\mid D'J)_\Omega=0$ on its output.

Three further applications concern the original systems under the
double Markov conditions.
\begin{itemize}
 \item If $\rho_{ABCD}$ is diagonal in product bases on $B,C,D$,
 the common algebra consists of functions of the conditional
 density operator of $A$. Its minimal central projections group
 triples $(b,c,d)$ of positive probability with identical
 conditional states of $A$. Either $BD$ or $CD$ determines the
 class label $J$. Recording it keeps the original $ABCD$
 marginal and gives $I(A;BC\mid DJ)=0$
 (Proposition~\ref{prop:diagonal-double-markov}).
 \item If $D$ is one-dimensional, the observations consist of
 the disjoint systems $B$ and $C$. The common algebra is
 commutative. Separate projective measurements on $B$ and $C$
 give the same state $\Omega_{ABCJ}$, with original $ABC$
 marginal and $I(A;BC\mid J)_\Omega=0$
 (Proposition~\ref{thm:tripartite-double-Markov}).
 \item If $\rho_{BCD}$ has full rank, its support is all of
 $\mathcal H_{BCD}$. The two regional operator images can then
 be compared on the full space, where their shared observables
 act on $D$. The common sufficient algebra therefore acts on
 the physical overlap $D$, and $I(A;BC\mid D)_\rho=0$
 (Proposition~\ref{thm:full-support-conditional-intersection}).
\end{itemize}

Double Markov statements enter subadditivity--doubling--rotation
arguments for Gaussian extremality and entropy inequalities
\cite{GengNair14,ajn19,HayashiZhao2026DoubleMarkov}.
Section~\ref{sec:preliminaries} develops the conditional-state
families and recovery, Section~\ref{sec:main-common-information}
proves the algebra and representation theorems, and
Section~\ref{sec:double-markov-applications} gives the applications.

Logarithms in entropy and relative entropy are base two. We write
$\mathcal L(\mathcal H)$ for bounded operators on $\mathcal H$.
We use finite-dimensional Hilbert spaces and finite classical registers.
\section{Preliminaries}\label{sec:preliminaries}
\label{sec:quantum-realization}
\subsection{Double Markov conditions and conditional states}

We work with finite-dimensional quantum systems $A,B,C$ with Hilbert spaces
$\mathcal H_A,\mathcal H_B,\mathcal H_C$.
A state on $A$ is a positive semidefinite operator $\rho_A$ with $\Tr\rho_A=1$;
for a composite system $AB$ we write $\rho_{AB}$ for a joint state and
$\rho_A=\Tr_B\rho_{AB}$ for the marginal.
To express Markov-type conditional independence in quantum terms, we use the conditional mutual
information (CMI) defined from von Neumann entropies, which are defined as
$S(A):=S(\rho_A):=-\Tr(\rho_A\log\rho_A)$.
We also write the conditional entropy as $S(A|B):=S(AB)-S(B)$.
For a tripartite state $\rho_{ABC}$, the CMI is
\begin{align}
I(A;C|B) := S(AB)+S(BC)-S(B)-S(ABC).
\end{align}
Strong subadditivity asserts $I(A;C|B)\ge 0$ for all $\rho_{ABC}$. The equality case
$I(A;C|B)=0$ is known to be equivalent to $\rho_{ABC}$ having a specific structural decomposition \cite{HJP}. 
Motivated by the classical identity ``$A-B-C$ is Markov $\Leftrightarrow I(A;C|B)=0$'',
we say that $A-B-C$ forms a (quantum) Markov chain if $I(A;C|B)=0$. 

\emph{Double Markov conditions.}
Now, we consider a four-party quantum system $ABCD$ with a joint
density operator $\rho_{ABCD}$. We say it satisfies the \textit{double Markov conditions} if
\begin{equation}
 I(A;C\mid BD)_\rho=0,\qquad I(A;B\mid CD)_\rho=0.
 \label{eq:quantum-crossed-conditions}
\end{equation}
In later discussions, we introduce the shorthand for the joint system $BCD$ as $Y$ and write $\rho_{AY}$ for the joint state on $A$ and $Y$, in other words,
\begin{align*}
    \mathcal H_Y:=\mathcal H_B\otimes \mathcal H_C\otimes \mathcal H_D,\qquad
    \rho_{AY}:=\rho_{ABCD}.
\end{align*}

We define the \textit{conditional states} for $Y$ to track the correlations with $A$. The conditional states compare the states on $Y$
conditioned on measurement outcomes on $A$. An effect $0\leq F\leq I_A$ represents one measurement outcome on $A$.
Its probability is $p_F:=\operatorname{Tr}(\rho_AF)$.
For $p_F>0$, the conditional state on $Y$ is
\begin{equation}
 \rho_{Y\mid F}:=\frac{\operatorname{Tr}_A\!\left[
 (F^{1/2}\otimes I_Y)\rho_{AY}(F^{1/2}\otimes I_Y)\right]}{p_F}.
 \label{eq:measurement-conditional-state}
\end{equation}
We consider these normalized density operators for
$F\in\mathcal F_+:=\{F:0\leq F\leq I_A,\ p_F>0\}$.
The outcome $F=I_A$ belongs to this set and gives
$\rho_{Y\mid I_A}=\rho_Y$.

Without loss of generality, we may assume $\mathcal H_A = \operatorname{supp}\rho_A$. However, the same assumption cannot be applied for $\rho_Y$, as it contains multiple systems $B,C,D$. We denote the support of $\rho_Y$ by $P_Y:=\operatorname{supp}\rho_Y$.

\subsection{Sufficient observations and recovery}
Let $\mathcal H_R$ be a Hilbert space.
An observation is a completely positive trace-preserving (CPTP) map
$\mathcal T:Y\to R$, acting on density operators.
Its observed conditional states are
$\rho_{R\mid F}:=\mathcal T(\rho_{Y\mid F})$.
We write $\mathcal T_{Y\to R}$ when the systems need to be specified.
It is \textit{sufficient} for the conditional states
\eqref{eq:measurement-conditional-state} if one recovery channel
$\mathcal R: R\to Y$ satisfies
\begin{equation}
 \mathcal R(\mathcal T(\rho_{Y\mid F}))=\rho_{Y\mid F}.
 \label{eq:steering-channel-sufficiency}
\end{equation}
for every outcome $F\in \mathcal F_+$.
Note that for each $F\in\mathcal F_+$ and observable $x$ on $Y$, we have
$p_F\operatorname{Tr}(\rho_{Y\mid F}x)
=\operatorname{Tr}[\rho_{AY}(F\otimes x)]$, and $\mathrm{span}\{\mathcal F_+\} = \mathcal L(\mathcal H_A)$.
Thus recovering every conditional state is equivalent to recovering
the joint state:
\begin{equation}
 (\operatorname{id}_A\otimes(\mathcal R\circ\mathcal T))(\rho_{AY})=\rho_{AY}.
 \label{eq:family-joint-recovery}
\end{equation}

Suppose the Hilbert spaces $\mathcal H_Y$ and $\mathcal H_R$ are such that $\mathcal H_Y = \mathcal H_R \otimes \mathcal H_Z$, then the observation channel $\mathcal T$ is naturally given by $\mathcal T = \operatorname{Tr}_Z$.
The conditional state on $R$ is
$\rho_{R\mid F}:=\operatorname{Tr}_Z\rho_{Y\mid F}$.
The next lemma supplies the characterization of quantum Markov chains.
\begin{lemma}[Quantum conditional independence and sufficiency]
\label{lem:quantum-CI-sufficiency}
For any finite-dimensional state $\rho_{ARZ}$, the condition
$I(A;Z\mid R)_\rho=0$ holds if and only if one channel
$\mathcal R: R\to Y$ satisfies
$\mathcal R(\rho_{R\mid F})=\rho_{RZ\mid F}$ for every $F\in\mathcal F_+$.
\end{lemma}

\begin{proof}
By the equality case of strong subadditivity
\cite{Petz1986SufficientSubalgebras,HJP}, $I(A;Z\mid R)_\rho=0$
if and only if there is a channel $\mathcal R: R\to Y$ such that
\[
 \rho_{ARZ}=(\operatorname{id}_A\otimes\mathcal R)(\rho_{AR}).
\]
Here,
\[
\mathcal R(s) = \rho_{Y}^{1/2}(\rho_R^{-1/2}s\rho_R^{-1/2}\otimes I_Z)\rho_{Y}^{1/2}.
\]
Combining with \eqref{eq:family-joint-recovery} proves the claim.
\end{proof}

\subsection{Sufficient algebras and minimal sufficient algebras}
Following
\cite[Section~1]{JencovaPetz2006Sufficiency}, we define
sufficient algebras through a map on observables.
Put
\[
 \mathcal N_Y:=\mathcal L(P_Y\mathcal H_Y),\qquad
 \phi_{Y\mid F}(x):=\Tr(\rho_{Y\mid F}x),
\]
and write
\[
 \mathsf S_Y:=
 \bigl(\mathcal N_Y,\{\phi_{Y\mid F}\}_{F\in\mathcal F_+}\bigr),
\]
We regard each conditional density operator as the state
$\phi_{Y\mid F}$ on the observable algebra $\mathcal N_Y$.

A unital $*$-subalgebra $\mathcal A\subseteq\mathcal N_Y$
contains $P_Y$ and is closed under linear combinations,
products, and adjoints. We say $\mathcal A$
is \emph{sufficient} for $\mathsf S_Y$ if there exists a
unital $2$-positive linear map
$\alpha_{\mathcal A}:\mathcal N_Y\to\mathcal A$ such that
\begin{equation}
 \phi_{Y\mid F}(\alpha_{\mathcal A}(x))=\phi_{Y\mid F}(x)
 \qquad
 (x\in\mathcal N_Y,\ F\in\mathcal F_+).
 \label{eq:sufficient-subalgebra-state}
\end{equation}

Here unitality means $\alpha_{\mathcal A}(P_Y)=P_Y$,
and $2$-positivity means that
$\operatorname{id}_{M_2}\otimes\alpha_{\mathcal A}$
preserves positivity. Since
$\rho_{Y\mid I_A}=\rho_Y$ is faithful, the map
can be chosen completely positive
\cite[Theorem~1]{JencovaPetz2006Sufficiency}.

We define the minimal sufficient algebra by
\begin{equation}
 \mathcal M(\mathsf S_Y)
 :=\bigcap_{\substack{
       \mathcal A\subseteq\mathcal N_Y\
          \text{ a unital }*\text{-subalgebra}\\
       \mathcal A\text{ sufficient for }\mathsf S_Y}}
       \mathcal A.
 \label{eq:minimal-sufficient-intersection}
\end{equation}
The sufficiency criterion in
\cite[Theorem~1]{JencovaPetz2006Sufficiency}
ensures that this intersection is itself sufficient.

\subsection{Relating sufficient observations and sufficient algebras}

For an observation $\mathcal T:Y\to R$, write
$\rho_R:=\mathcal T(\rho_Y)$ and take $P_R\mathcal H_R$
to be its support space. This contains the supports of all
observed conditional states, since
\(
 \rho_{R\mid F}\leq p_F^{-1}\rho_R.
\)

The trace adjoint
$\mathcal T^*:\mathcal N_R\to\mathcal N_Y$ is defined by
\[
 \Tr[\mathcal T(t)b]=\Tr[t\mathcal T^*(b)].
\]
It is unital and completely positive. The observed experiment
is $\mathsf S_R=(\mathcal N_R,\{\phi_{R\mid F}\}_{F\in\mathcal F_+})$,
where
\[
 \phi_{R\mid F}(b):=\Tr(\rho_{R\mid F}b)
 =\phi_{Y\mid F}(\mathcal T^*(b)).
\]

We use $\mathcal R$ for the recovery channel, and put
$\mathcal N_R:=\mathcal L(P_R\mathcal H_R)$. Similarly, a recovery channel has a unital completely positive
adjoint $\mathcal R^*:\mathcal N_Y\to\mathcal N_R$.
By trace duality, the recovery condition
\eqref{eq:steering-channel-sufficiency} is equivalent to
\begin{equation}
 \phi_{Y\mid F}\!\left(\mathcal T^*(\mathcal R^*(x))\right)
 =\phi_{Y\mid F}(x)
 \qquad
 (x\in\mathcal N_Y,\ F\in\mathcal F_+).
 \label{eq:sufficient-observation-observable}
\end{equation}

A natural question to ask is whether $\mathcal T^*(\mathcal N_R)$ is a sufficient algebra for $\mathsf S_Y$. The answer is almost yes, the only problem is that $\mathcal T^*(\mathcal N_R)$ may not be a $*$-subalgebra. The key to resolve this problem is the multiplicative domain of $\mathcal T^*$, which is a $*$-subalgebra of $\mathcal N_R$.

For an observation $\mathcal T:Y\to R$, the
\emph{multiplicative domain} of its unital completely positive
adjoint $\mathcal T^*$ is the set
$\mathcal N_{\mathcal T^*}$ of all $b\in\mathcal N_R$
satisfying
\[
\begin{aligned}
 \mathcal T^*(x^*x)
 &=\mathcal T^*(x)^*\mathcal T^*(x),\quad
 \mathcal T^*(xx^*)
 =\mathcal T^*(x)\mathcal T^*(x)^*.
\end{aligned}
\]
It is a unital $*$-subalgebra of $\mathcal N_R$, and
\[
\begin{aligned}
 \mathcal T^*(xy)&=\mathcal T^*(x)\mathcal T^*(y),\quad
 \mathcal T^*(yx)=\mathcal T^*(y)\mathcal T^*(x)
\end{aligned}
\]
for every $x\in\mathcal N_{\mathcal T^*}$ and
$y\in\mathcal N_R$
\cite[Lemma~5]{JencovaPetz2006Sufficiency}.
Thus $\mathcal T^*$ restricts to a unital $*$-homomorphism
on its multiplicative domain. In particular, if $x$ and
$\mathcal T^*(x)$ are unitary in their respective supported
algebras, then $x\in\mathcal N_{\mathcal T^*}$.
If $\mathcal T$ is sufficient for $\mathsf S_Y$, then
\[
 \mathcal T^*\bigl(\mathcal N_{\mathcal T^*}\bigr)
\]
is a sufficient subalgebra of $\mathcal N_Y$
\cite[Theorem~2]{JencovaPetz2006Sufficiency}.

Now, we consider $\mathcal H_Y = \mathcal H_R \otimes \mathcal H_Z$ with $\mathcal T=\Tr_Z$. Then $\mathcal T^*:\mathcal N_R\to\mathcal N_Y$ is given by $$\mathcal T^*(x)= P_Y (x\otimes I_Z) P_Y,$$
while
$$\mathcal R^*(x) = \rho_{R}^{-1/2}\Tr_Z(\rho_Y^{1/2}x\rho_Y^{1/2})\rho_{R}^{-1/2}.$$

Combining with the Lemma \ref{lem:quantum-CI-sufficiency}, we summarize the equivalence between the double Markov condition and the sufficient observation and sufficient subalgebra as follows.
\begin{proposition}\label{three-equivalence}
    For any finite-dimensional state $\rho_{ARZ}$, let $\mathcal T$ be $\Tr_Z$, and conditional states $\{\rho_{Y|F}\}_{F\in\mathcal F_+}$, experiment $\mathsf S_Y$ be defined accordingly. Then the following are equivalent:
\begin{enumerate}
    \item $I(A;Z\mid R)_\rho=0$.
    \item $\mathcal T$ is sufficient observation for $\{\rho_{Y|F}\}_{F\in\mathcal F_+}$.
    \item $\mathcal T^*(\mathcal N_{\mathcal T^*})$ is sufficient subalgebra for $\mathsf S_Y$.
\end{enumerate}
\end{proposition}

The next lemma shows that sufficient subalgebras are invariant when we assume $\rho_{AY}$ has full support on $\mathcal H_A\otimes P_Y\mathcal H_Y$.

\begin{lemma}\label{lem:common-support-regularization}
    For any $\rho_{AY}$, put
\begin{equation}
 \widetilde\rho_{AY}:=\tfrac12(\rho_{AY}+\rho_A\otimes\rho_Y).
 \label{eq:physical-state-mixing}
\end{equation}
For any channel $\Lambda$ on $Y$, it holds that
\begin{align}
    (\operatorname{id}_A\otimes\Lambda)(\rho_{AY})=\rho_{AY} \quad
     \iff \quad &(\operatorname{id}_A\otimes\Lambda)(\widetilde\rho_{AY})=\widetilde\rho_{AY}.
\end{align}
Consequently, the sufficient subalgebras for $\mathsf S_Y$ and $\widetilde{\mathsf S}_Y$ are the same, where 
\begin{align*}
    &\widetilde{\mathsf S}_Y:=(\mathcal N_Y,\{\widetilde\phi_{Y\mid F}\}_{F\in\mathcal F_+}),\\
    &\widetilde\phi_{Y\mid F}(x):=\Tr(\widetilde\rho_{Y\mid F}x),\\
    &\widetilde\rho_{Y\mid F}:=\frac{\operatorname{Tr}_A[(F^{1/2}\otimes I_Y)\widetilde\rho_{AY}(F^{1/2}\otimes I_Y)]}{p_F}.
\end{align*}

\end{lemma}

\begin{proof}
Taking partial traces in \eqref{eq:physical-state-mixing} gives
\[
 \widetilde\rho_A=\rho_A,\qquad
 \widetilde\rho_Y=\rho_Y.
\]
If $\operatorname{id}_A\otimes\Lambda$ fixes either joint state,
taking the partial trace over $A$ gives $\Lambda(\rho_Y)=\rho_Y$.
Thus it fixes $\rho_A\otimes\rho_Y$. The equivalence follows from
\[
 \widetilde\rho_{AY}
 =\frac12(\rho_{AY}+\rho_A\otimes\rho_Y),
 \qquad
 \rho_{AY}=2\widetilde\rho_{AY}-\rho_A\otimes\rho_Y.
\]

To show the sufficient algebras are identical, it suffices that for each $\alpha$, 
\[
 \widetilde\phi_{Y\mid F}\circ\alpha
   =\widetilde\phi_{Y\mid F}\quad \forall F
 \quad\Longleftrightarrow\quad
 \phi_{Y\mid F}\circ\alpha
   =\phi_{Y\mid F}\quad \forall F.
\]
Note that the conditional
states satisfy
\[
 \widetilde\rho_{Y\mid F}
 =\frac12(\rho_{Y\mid F}+\rho_Y),
 \qquad
 \widetilde\rho_{Y\mid I_A}=\rho_Y.
\]
If $\alpha$ preserves the regularized family, then $F=I_A$
shows that it preserves the reference state $\rho_Y$. Subtracting
the reference state in the mixing formula gives preservation of
each original conditional state. Conversely, a map preserving the
original family also preserves $\rho_Y$ and hence preserves each
regularized state by linearity.
\end{proof}

\begin{lemma}\label{lem:generated-algebra}
     Let $W^*$ denote the generated unital *-subalgebra. Then the minimal sufficient algebra $\mathcal M(\mathsf S_Y)$ is given by
    \begin{equation}
 \mathcal M(\mathsf S_Y)=W^*\{
  \rho_{Y|F}^{it}\rho_Y^{-it}:F\in\mathcal F_+,\ t\in\mathbb R\}.
 \label{eq:quantum-minimal-generators}
\end{equation}
\end{lemma}

\begin{proof}
The reference state $\phi_{Y\mid I_A}$ has faithful density
$\rho_Y$ on $P_Y\mathcal H_Y$. The finite-dimensional relative
Connes cocycles are therefore
\[
 [D\phi_{Y\mid F},D\phi_{Y\mid I_A}]_t
 =\rho_{Y\mid F}^{it}\rho_Y^{-it}.
\]
By \cite[Theorem~1]{JencovaPetz2006Sufficiency}, a unital
$*$-subalgebra $\mathcal A$ is sufficient if and only if it contains $\rho_{Y\mid F}^{it}\rho_Y^{-it}$ for every $F\in\mathcal F_+$ and $t\in\mathbb R$. The result follows.
\end{proof}
\begin{remark}
    The above lemma holds for $\mathcal M({\mathsf S}_R)$ as well. In general, $W^*$ denotes the generated von Neumann algebra. In finite dimensions, the von Neumann algebra is the same as the unital *-subalgebra.
\end{remark}

\section{Common sufficient algebra and its isometric reduction to subsystems}
\label{sec:main-common-information}

Throughout this section, assume the double Markov conditions
\eqref{eq:quantum-crossed-conditions}. 
For $R=BD,CD$ (thus $Z = C,B$), let $P_R$ be the support projection of $\rho_R$, and let
\[
 \mathcal T_{BD}(\rho)=\Tr_C \rho,\qquad
 \mathcal T_{CD}(\rho)=\Tr_B \rho,
\]
restricted to the marginal supports.

\subsection{Common sufficient algebra}
\label{subsec:common-sufficient-algebra}

The two adjoints compare regional observables on the support of
$\rho_Y=\rho_{BCD}$. For $R=BD$ take $Z=C$, and for $R=CD$ take
$Z=B$. Extend $x\in\mathcal N_R=\mathcal L(P_R\mathcal H_R)$ by
zero on the orthogonal complement of $P_R\mathcal H_R$ in
$\mathcal H_R$. Form $x\otimes I_Z$ in tensor order $R,Z$ and
view it in the common tensor order $B,C,D$. Project onto the
support of $\rho_Y$ on both sides. With $P_Y$ the support
projection, the supported adjoint is
\[
 \mathcal T_R^*(x)=P_Y(x\otimes I_Z)P_Y,
 \qquad x\in\mathcal N_R.
\]

Let $\mathsf S_{BD}$ and $\mathsf S_{CD}$ denote the observed
experiments defined from the same effects on $A$.

\begin{theorem}[Common sufficient algebra]
\label{thm:main-common-information}
Let $\rho_{ABCD}$ satisfy
\eqref{eq:quantum-crossed-conditions}, and put $Y=BCD$.
Then
\begin{equation}
 \begin{aligned}
 \mathcal M_{\mathrm{com}}
 &:=\mathcal M(\mathsf S_Y)\\
 &=\mathcal T_{BD}^*
       \bigl(\mathcal M(\mathsf S_{BD})\bigr)\\
 &=\mathcal T_{CD}^*
       \bigl(\mathcal M(\mathsf S_{CD})\bigr).
 \end{aligned}
 \label{eq:quantum-common-images}
\end{equation}
For each $R=BD,CD$, the restriction
\[
 \mathcal T_R^*|_{\mathcal M(\mathsf S_R)}:
 \mathcal M(\mathsf S_R)\longrightarrow\mathcal M_{\mathrm{com}}
\]
is a $*$-isomorphism.

There exists a conditional expectation
$\mathbb E_Y:\mathcal N_Y\to\mathcal N_Y$ with range
$\mathcal M_{\mathrm{com}}$ satisfying
\[
 \phi_{Y\mid F}\circ \mathbb E_Y=\phi_{Y\mid F}
 \qquad(F\in\mathcal F_+).
\]
Its trace-adjoint channel $\mathbb E^*_Y$ preserves the joint state:
\begin{equation}
 (\operatorname{id}_A\otimes\mathbb E^*_Y)(\rho_{ABCD})
 =\rho_{ABCD}.
 \label{eq:common-algebra-joint-state-preservation}
\end{equation}
\end{theorem}

\begin{proof}
By Proposition~\ref{three-equivalence}, each partial trace
$\mathcal T_R$, for $R=BD,CD$, has a recovery channel $\mathcal R_R$
such that $\operatorname{id}_A\otimes(\mathcal R_R\circ\mathcal T_R)$
fixes $\rho_{AY}$. By Lemma~\ref{lem:common-support-regularization},
the same composition fixes $\widetilde\rho_{AY}$, so $\mathcal T_R$
is also sufficient for the regularized family.

The reference states $\rho_Y$ and $\rho_R=\mathcal T_R(\rho_Y)$
and their supports are unchanged by regularization. The observed
conditional states satisfy
\[
 \widetilde\rho_{R\mid F}
 =\mathcal T_R(\widetilde\rho_{Y\mid F})
 =\tfrac12(\rho_{R\mid F}+\rho_R).
\]
On each of $Y$ and $R$, the regularized conditional states are
bounded below by half the reference state, so they are faithful on
the corresponding support. Lemma~\ref{lem:common-support-regularization}
also gives the same minimal sufficient algebra before and after
mixing on $Y$. Applied to $\rho_{AR}$, it gives the same conclusion
on $R$.

Now fix $R=BD$ or $CD$. Sufficiency of $\mathcal T_R$ for the
regularized family gives the cocycle identity
\cite[Theorem~2, (ii)$\Rightarrow$(iv)]{JencovaPetz2006Sufficiency}
\[
 \mathcal T_R^*
 \bigl(\widetilde\rho_{R\mid F}^{it}\rho_R^{-it}\bigr)
 =\widetilde\rho_{Y\mid F}^{it}\rho_Y^{-it}
\]
for every $F\in\mathcal F_+$ and $t\in\mathbb R$.
By Lemma~\ref{lem:generated-algebra}, the cocycles on the left generate
$\mathcal M(\mathsf S_R)$, and those on the right generate
$\mathcal M(\mathsf S_Y)$. Fix a generator
$v=\widetilde\rho_{R\mid F}^{it}\rho_R^{-it}$. Both $v$ and its
image are unitary on their respective supports, and
$\mathcal T_R^*(P_R)=P_Y$. The two equalities
\begin{align*}
 \mathcal T_R^*(v^*v)
 &=P_Y=\mathcal T_R^*(v)^*\mathcal T_R^*(v),\\
 \mathcal T_R^*(vv^*)
 &=P_Y=\mathcal T_R^*(v)\mathcal T_R^*(v)^*
\end{align*}
place $v$ in the multiplicative domain of $\mathcal T_R^*$.
This domain is a $*$-subalgebra and contains every generator of
$\mathcal M(\mathsf S_R)$. Hence the restriction of $\mathcal T_R^*$
to $\mathcal M(\mathsf S_R)$ is a unital $*$-homomorphism.

To prove injectivity, let $x\in\mathcal M(\mathsf S_R)$ satisfy
$\mathcal T_R^*(x)=0$. Then
\[
 \begin{aligned}
 \Tr(\rho_Rx^*x)
 &=\Tr[\rho_Y\mathcal T_R^*(x^*x)]\\
 &=\Tr[\rho_Y\mathcal T_R^*(x)^*\mathcal T_R^*(x)]
 =0.
 \end{aligned}
\]
Faithfulness of $\rho_R$ implies $x=0$.
Thus the restriction is a $*$-isomorphism onto its image.

We now identify this image. A $*$-isomorphism maps a generated
algebra onto the algebra generated by the images of its generators.
Hence, with $F\in\mathcal F_+$ and $t\in\mathbb R$,
\[
 \begin{aligned}
 \mathcal T_R^*\bigl(\mathcal M(\mathsf S_R)\bigr)
 &=W^*\!\left\{
   \mathcal T_R^*
   \bigl(\widetilde\rho_{R\mid F}^{it}\rho_R^{-it}\bigr)
   \right\}\\
 &=W^*\!\left\{\widetilde\rho_{Y\mid F}^{it}\rho_Y^{-it}\right\}\\
 &=\mathcal M(\mathsf S_Y),
 \end{aligned}
\]
where the last equality is Lemma~\ref{lem:generated-algebra}.
Applying this to both partial traces proves
\eqref{eq:quantum-common-images} and the asserted
$*$-isomorphisms.

For the regularized family on $Y$, the sufficient-subalgebra
criterion \cite[Theorem~1, (v)]{JencovaPetz2006Sufficiency} gives a
conditional expectation $\mathbb E_Y$ with range
$\mathcal M_{\mathrm{com}}$ that preserves every
$\widetilde\phi_{Y\mid F}$. Its trace adjoint
$\mathbb E_Y^*$ therefore preserves every $\widetilde\rho_{Y\mid F}$.
By \eqref{eq:family-joint-recovery}, applied to the regularized family,
\[
 (\operatorname{id}_A\otimes\mathbb E_Y^*)(\widetilde\rho_{AY})
 =\widetilde\rho_{AY}.
\]
Lemma~\ref{lem:common-support-regularization} now gives
\eqref{eq:common-algebra-joint-state-preservation} for the original
state. Testing this identity against the effects $F$ on $A$ gives
$\mathbb E_Y^*(\rho_{Y\mid F})=\rho_{Y\mid F}$ for all
$F\in\mathcal F_+$. Thus $\phi_{Y\mid F}\circ\mathbb E_Y=\phi_{Y\mid F}$
by trace duality.

\end{proof}

\subsection{Isometric reduction on a sufficient subsystem}
\label{subsec:isometric-reduction}

We first identify the common sufficient algebra in the
Koashi--Imoto decomposition of $\rho_{AY}$. We then construct
an isometry whose output contains a subsystem on which this algebra
acts.

The minimal Koashi--Imoto decomposition
\cite{KoashiImoto2002} associated with the family
$\{\rho_{Y\mid F}\}_{F\in\mathcal F_+}$ is unique up to unitary
equivalence and permutation of blocks. It decomposes the support
$P_Y\mathcal H_Y$ as follows.

\begin{equation}
 \begin{aligned}
 P_Y\mathcal H_Y
 &=\bigoplus_j(L_j\otimes R_j),\\
 \rho_{AY}
 &=\bigoplus_j p_j\rho_{AL_j}\otimes\tau_j.
 \end{aligned}
 \label{eq:minimal-KI-block-form}
\end{equation}

\begin{definition}[Minimal Koashi--Imoto decomposition]
Consider \eqref{eq:minimal-KI-block-form}, with $p_j>0$,
$\sum_jp_j=1$, and normalized states $\rho_{AL_j}$ and $\tau_j$.
The decomposition is called \emph{minimal} if the following
conditions hold \cite[Definition~1]{KoashiImoto2002}.

\begin{enumerate}
\item 
For every $j$, the only orthogonal projections
$P\in\mathcal L(L_j)$ satisfying
\[
 (I_A\otimes P)\rho_{AL_j}
 =\rho_{AL_j}(I_A\otimes P)
\]
are $P=0$ and $P=I_{L_j}$.
Equivalently, $L_j$ admits no nontrivial orthogonal direct-sum
decomposition with respect to which $\rho_{AL_j}$ is block
diagonal.

\item 
For $j\ne j'$, there is no unitary $V:L_j\to L_{j'}$ such that
\[
 (I_A\otimes V)\rho_{AL_j}(I_A\otimes V^*)
 =\rho_{AL_{j'}}.
\]
\end{enumerate}

\end{definition}

In the original work by Koashi and Imoto, it was called the "maximal decomposition".

We first show that
\begin{align*}
  \mathcal M(\mathsf S_Y) &= \bigoplus_j\bigl(\mathcal L(L_j)\otimes I_{R_j}\bigr).
\end{align*}

Let $\Pi_j$ be the projection onto $L_j\otimes R_j$ in the
minimal Koashi--Imoto decomposition \eqref{eq:minimal-KI-block-form}.
For every $F\in\mathcal F_+$, \eqref{eq:measurement-conditional-state}
gives
\[
 \rho_{Y\mid F}
 =\bigoplus_j \frac{p_j}{p_F}
   \Tr_A\!\bigl[
     (F^{1/2}\otimes I_{L_j})\rho_{AL_j}
     (F^{1/2}\otimes I_{L_j})
   \bigr]\otimes\tau_j.
\]
Taking $F=I_A$ gives
\[
 \rho_Y=\bigoplus_j p_j(\Tr_A\rho_{AL_j})\otimes\tau_j.
\]
Since $\rho_Y$ is faithful on $P_Y\mathcal H_Y$, each $\tau_j$
is faithful on $R_j$. In block $j$, both $\rho_{Y\mid F}$ and
$\rho_Y$ have the right factor $\tau_j$, so the same is true of
$\widetilde\rho_{Y\mid F}=\tfrac12(\rho_{Y\mid F}+\rho_Y)$.
The regularized conditional states are faithful on
$P_Y\mathcal H_Y$. Thus the right factors of the regularized
cocycles $\widetilde\rho_{Y\mid F}^{it}\rho_Y^{-it}$ cancel as
$\tau_j^{it}\tau_j^{-it}=I_{R_j}$. By
Lemmas~\ref{lem:common-support-regularization}
and \ref{lem:generated-algebra},
\[
 \mathcal M(\mathsf S_Y)
 \subseteq\bigoplus_j\bigl(\mathcal L(L_j)\otimes I_{R_j}\bigr).
\]

For the reverse inclusion, take the conditional expectation
$\mathbb E_Y$ from Theorem~\ref{thm:main-common-information}.
Since $\phi_{Y\mid F}\circ\mathbb E_Y=\phi_{Y\mid F}$, its trace
adjoint satisfies
\[
 \mathbb E_Y^*(\rho_{Y\mid F})=\rho_{Y\mid F}
 \qquad(F\in\mathcal F_+).
\]
The given decomposition is minimal for this conditional-state family.
Indeed, since $\mathcal F_+$ spans $\mathcal L(\mathcal H_A)$, a
projection commuting with all weighted left blocks for a fixed $j$
gives a projection $P$ satisfying
$[I_A\otimes P,\rho_{AL_j}]=0$. Likewise, if a unitary identified
the weighted left blocks for $j$ with a fixed multiple of those for
$j'$, taking traces at $F=I_A$ would give the multiple $p_j/p_{j'}$.
Testing against all $F$ would then give the unitary equivalence
between $\rho_{AL_j}$ and $\rho_{AL_{j'}}$ excluded by minimality.

The Koashi--Imoto preservation theorem
\cite[Theorem~3]{KoashiImoto2002} therefore gives a Kraus
representation of $\mathbb E_Y^*$ with all Kraus operators in
$\bigoplus_j(I_{L_j}\otimes\mathcal L(R_j))$. Every
$x=\bigoplus_j(x_j\otimes I_{R_j})$, with $x_j\in\mathcal L(L_j)$,
commutes with these operators and their adjoints. Hence
\[
 \mathbb E_Y(x)=x\,\mathbb E_Y(P_Y)=x.
\]
Thus $x$ belongs to the range of $\mathbb E_Y$, which proves
\[
 \mathcal M_{\mathrm{com}}
 =\mathcal M(\mathsf S_Y)
 =\bigoplus_j\bigl(\mathcal L(L_j)\otimes I_{R_j}\bigr).
\]
Its center is $\bigoplus_j\mathbb C\Pi_j$, so the $\Pi_j$ are
precisely its minimal central projections.

We now construct an isometry from $P_Y\mathcal H_Y$ to
$EFJ$. The register $J$ records the block index, $F$ holds the
left factor of that block, and $E$ holds its right factor. The
representation of $\mathcal M_{\mathrm{com}}$ will act on $FJ$.

Choose
\[
 \dim\mathcal H_E=\max_j\dim R_j,
 \qquad
 \dim\mathcal H_F=\max_j\dim L_j,
\]
and isometries $W_j:R_j\to\mathcal H_E$ and
$V_j:L_j\to\mathcal H_F$. Define
$U:P_Y\mathcal H_Y\to\mathcal H_{EFJ}$ blockwise by
\[
 U(\xi\otimes\eta)
 =W_j\eta\otimes V_j\xi\otimes |j\rangle_J,
 \qquad \xi\in L_j,\quad \eta\in R_j.
\]
Each block map preserves inner products, and the vectors
$|j\rangle_J$ make the images of distinct blocks orthogonal.
Thus $U$ is an isometry.

Let $\rho_{AEFJ}:=(I_A\otimes U)\rho_{AY}(I_A\otimes U^*)$.
Writing the tensor factors in the order $A,F,E,J$ gives
\begin{align*}
 \rho_{AFEJ}
 &=\sum_j p_j
   \bigl[(I_A\otimes V_j)\rho_{AL_j}(I_A\otimes V_j^*)\bigr]\\
 &\qquad\otimes (W_j\tau_j W_j^*)\otimes |j\rangle\langle j|_J .
\end{align*}
The joint output is classical on $J$. Conditioned on $J=j$,
the state on $AF$ is in product with the state on $E$. Hence
\[
 I(A;E\mid FJ)_\rho
 =\sum_j p_j I(A;E\mid F)_{\rho_{AFE\mid J=j}}
 =0.
\]

For $m=\bigoplus_j(m_j\otimes I_{R_j})\in\mathcal M_{\mathrm{com}}$,
define
\[
 \pi(m)=\sum_j V_jm_jV_j^*\otimes |j\rangle\langle j|_J .
\]
This is a faithful $*$-representation with unit
$\pi(P_Y)=\operatorname{supp}\rho_{FJ}$. On the $j$-th input block,
both $Um$ and $(I_E\otimes\pi(m))U$ send $\xi\otimes\eta$ to
$W_j\eta\otimes V_jm_j\xi\otimes|j\rangle_J$. Thus
\[
 Um=(I_E\otimes\pi(m))U
 \qquad(m\in\mathcal M_{\mathrm{com}}).
\]

Motivated by this construction, we now state the main theorem of this section. Regarding the state $\Omega$ induced by $U$, one should note that although in later applications, $U$ preserves the marginals as $\Tr_J \Omega_{ABCDJ} = \rho_{ABCD}$, it is not required in the following theorem. A more precise way is to write the output systems as $A', B', C', D', J$, since it causes no ambiguity once noted by the reader, we stick to the notation $A,B,C,D,J$.

\begin{theorem}[Representation of sufficient information]
\label{thm:classical-quantum-realization}
For a fixed finite register $J$ and isometry
$U:P_Y\mathcal H_Y\to\mathcal H_{BCJD}$,
write $\Omega:=(I_A\otimes U)\rho_{AY}(I_A\otimes U^*)$.
Assume this joint output is classical on $J$. The following two conditions
are equivalent:
\begin{enumerate}
\item\label{cond:JD-screening} The output satisfies
\begin{equation}
 I(A;BC\mid JD)_\Omega=0.
 \label{eq:JD-output-screening}
\end{equation}
\item\label{cond:JD-compatible} There is an injective $*$-representation
\[\pi_U:\mathcal M_{\mathrm{com}}\to P_{JD}(\mathcal D_J\otimes\mathcal L(\mathcal H_D))P_{JD}\]
with $\pi_U(P_Y)=P_{JD}$,
such that
\begin{equation}
 Um=(I_{BC}\otimes\pi_U(m))U
 \quad(m\in\mathcal M_{\mathrm{com}}).
 \label{eq:JD-compatible-realization}
\end{equation}
Here $P_{JD}:=\operatorname{supp}\Omega_{JD}$, $\mathcal D_J$ is the diagonal algebra.
\end{enumerate}

Allowing $J$ and $U$ to vary, the following assertions are equivalent:
\begin{enumerate}
\renewcommand{\theenumi}{\alph{enumi}}
\renewcommand{\labelenumi}{\theenumi)}
\item\label{assert:JD-exists} There are a finite register $J$ and an isometry
$U:P_Y\mathcal H_Y\to\mathcal H_{B}\otimes
\mathcal H_{C}\otimes\mathcal H_J\otimes\mathcal H_{D}$
whose output $\Omega$ is classical on $J$ and satisfies
\eqref{eq:JD-output-screening}.
\item\label{assert:JD-embedding} For some finite $J$, there is an injective
$*$-homomorphism
$\pi:\mathcal M_{\mathrm{com}}\to\mathcal D_J\,\otimes \mathcal L(\mathcal H_D)$,
which is unital on its support $\pi(P_Y)\mathcal H_{DJ}$.
\item\label{assert:JD-dimension} Every $L_j$ satisfies
$\dim L_j\leq\dim\mathcal H_D$.
\end{enumerate}
\end{theorem}

\begin{proof}
For the fixed isometry $U$, the observation of $JD$ is
\[
 \mathcal T(t):=\Tr_{BC}(UtU^*).
\]
On the occupied spaces its adjoint is
$\mathcal T^*(b)=U^*(I_{BC}\otimes b)U$, where $b$ is extended
by zero outside $P_{JD}\mathcal H_{JD}$.

Assume \eqref{eq:JD-compatible-realization}. Multiplying the
intertwining relation by $U^*$ on the left gives
$\mathcal T^*(\pi_U(m))=m$ for $m\in\mathcal M_{\mathrm{com}}$.
Since $\pi_U(P_Y)=P_{JD}$, the map
$\mathcal R^*:=\pi_U\circ\mathbb E_Y$ is unital and completely
positive on the occupied spaces, so it defines a recovery channel
$\mathcal R$. Moreover,
\[
 (\mathcal R\circ\mathcal T)^*
 =\mathcal T^*\circ\pi_U\circ\mathbb E_Y
 =\mathbb E_Y.
\]
By \eqref{eq:common-algebra-joint-state-preservation},
$\mathcal R\circ\mathcal T$ preserves $\rho_{AY}$ when applied
on $Y$. Thus the channel $s\mapsto U\mathcal R(s)U^*$ recovers
$\Omega$ from $\Omega_{AJD}$. Lemma~\ref{lem:quantum-CI-sufficiency}
gives \eqref{eq:JD-output-screening}.

Conversely, suppose \eqref{eq:JD-output-screening} holds.
Lemma~\ref{lem:quantum-CI-sufficiency} gives a recovery channel
$\mathcal Q$ from $JD$ to $BCJD$ with
$(\operatorname{id}_A\otimes\mathcal Q)(\Omega_{AJD})=\Omega$.
In particular, $\mathcal Q(\Omega_{JD})=U\rho_Y U^*$. For any positive
operator $s$ on $P_{JD}\mathcal H_{JD}$, faithfulness of $\Omega_{JD}$
on that space gives $0\leq s\leq c\,\Omega_{JD}$ for some $c\geq 0$.
Positivity of $\mathcal Q$ therefore gives
\[
 0\leq\mathcal Q(s)\leq c\,U\rho_YU^*.
\]
Thus $\mathcal Q(s)$ is supported on the range of $U$.
Consequently, $\mathcal R(s):=U^*\mathcal Q(s)U$ defines a completely
positive map from $\mathcal L(P_{JD}\mathcal H_{JD})$ to $\mathcal N_Y$.
It is trace preserving because $\mathcal Q$ preserves the trace and
$\Tr(U^*\mathcal Q(s)U)=\Tr(\mathcal Q(s)UU^*)=\Tr(\mathcal Q(s))$
on this support. Applying $\operatorname{id}_A\otimes\mathcal R$
to $\Omega_{AJD}$ gives $\rho_{AY}$. Since
$\Omega_{AJD}=(\operatorname{id}_A\otimes\mathcal T)(\rho_{AY})$,
this proves that $\mathcal T$ is sufficient.

We now identify the minimal sufficient algebra on the observed
system $JD$. Put $\widetilde\Omega_{JD\mid F}
:=\mathcal T(\widetilde\rho_{Y\mid F})$.
Lemma~\ref{lem:common-support-regularization} preserves sufficiency
of $\mathcal T$ and the minimal sufficient algebras on both systems.
The regularized conditional states are faithful on the supports
$P_Y\mathcal H_Y$ and $P_{JD}\mathcal H_{JD}$, respectively,
because each dominates half of its reference state.
Sufficiency of $\mathcal T$ therefore gives
\cite[Theorem~2, (ii)$\Rightarrow$(iv)]{JencovaPetz2006Sufficiency}
\[
 \mathcal T^*\bigl(\widetilde\Omega_{JD\mid F}^{it}
                   \Omega_{JD}^{-it}\bigr)
 =\widetilde\rho_{Y\mid F}^{it}\rho_Y^{-it}
 \quad(F\in\mathcal F_+,\ t\in\mathbb R).
\]
As in the proof of Theorem~\ref{thm:main-common-information},
each observed cocycle and its image are unitary on their
respective supports. Thus the observed cocycles lie in the
multiplicative domain of $\mathcal T^*$. By
Lemma~\ref{lem:generated-algebra}, they generate
$\mathcal M(\mathsf S_{JD})$, and their images generate
$\mathcal M_{\mathrm{com}}$. The restriction of $\mathcal T^*$ is
therefore a $*$-homomorphism onto $\mathcal M_{\mathrm{com}}$.
If $\mathcal T^*(x)=0$ for $x\in\mathcal M(\mathsf S_{JD})$, then
\begin{align*}
 \Tr(\Omega_{JD}x^*x)
 &\overset{(a)}{=}\Tr[\rho_Y\mathcal T^*(x^*x)]\\
 &\overset{(b)}{=}\Tr[\rho_Y\mathcal T^*(x)^*\mathcal T^*(x)]
 =0.
\end{align*}
Equality (a) uses $\Omega_{JD}=\mathcal T(\rho_Y)$ and trace duality.
Equality (b) uses that $\mathcal T^*$ is a $*$-homomorphism on
$\mathcal M(\mathsf S_{JD})$. Faithfulness of $\Omega_{JD}$ on its
support implies $x=0$.
This proves that
\[
 \mathcal T^*|_{\mathcal M(\mathsf S_{JD})}:
 \mathcal M(\mathsf S_{JD})\longrightarrow\mathcal M_{\mathrm{com}}
\]
is a $*$-isomorphism. Let $\pi_U$ be its inverse.
Joint classicality on $J$ makes all the regularized
observed conditional states and their reference state
$\Omega_{JD}$ block diagonal on $J$. Their cocycles are therefore
block diagonal on $J$ as well. By Lemma~\ref{lem:generated-algebra},
\[
 \mathcal M(\mathsf S_{JD})
 \subseteq
 P_{JD}\bigl(\mathcal D_J\otimes\mathcal L(\mathcal H_D)\bigr)P_{JD}.
\]
The unit of $\pi_U$ is $P_{JD}$. For $m\in\mathcal M_{\mathrm{com}}$
and $\xi\in P_Y\mathcal H_Y$, the identity
$\mathcal T^*(\pi_U(m))=m$ and multiplicativity of $\pi_U$ give
\begin{align*}
 &\bigl\|(I_{BC}\otimes\pi_U(m))U\xi-Um\xi\bigr\|^2\\
 &\quad=
 \bigl\langle\xi,
 [\mathcal T^*(\pi_U(m^*m))-m^*m]\xi\bigr\rangle
 =0.
\end{align*}
This proves \eqref{eq:JD-compatible-realization}.

For the existence assertions, \ref{assert:JD-exists}) implies
\ref{assert:JD-embedding}) by the fixed-isometry equivalence.
The representation $\pi_U$ takes values in the corner supported
on $P_{JD}$. Extending its operators by zero on the orthogonal
complement gives the embedding into
$\mathcal D_J\otimes\mathcal L(\mathcal H_D)$ in
\ref{assert:JD-embedding}).

Suppose \ref{assert:JD-embedding}) holds. Injectivity ensures that
each simple summand $\mathcal L(L_j)$ of $\mathcal M_{\mathrm{com}}$
has a nonzero representation in some diagonal block of $J$.
In that block the representation is injective by simplicity, so
the images of $\dim L_j$ mutually orthogonal rank-one projections
on $L_j$ are nonzero orthogonal projections on $\mathcal H_D$.
Thus $\dim L_j\leq\dim\mathcal H_D$ for every $j$, proving
\ref{assert:JD-dimension}).

Finally, assume \ref{assert:JD-dimension}).
Choose isometries $V_j:L_j\to\mathcal H_D$ and diagonalize
$\tau_j=\sum_r\lambda_{jr}|r\rangle\langle r|$ in an orthonormal
basis of $R_j$. Let $J$ have an orthonormal basis
$\{|j,r\rangle_J\}_{j,r}$ and choose a unit vector $|0\rangle_{BC}$.
In the block coordinates of \eqref{eq:minimal-KI-block-form},
define
\[
 U(\xi\otimes|r\rangle)
 =|0\rangle_{BC}\otimes|j,r\rangle_J\otimes V_j\xi,
 \qquad \xi\in L_j.
\]
The orthogonal labels $(j,r)$ and the isometries $V_j$ make $U$
an isometry. The input is block diagonal in $j$, and each $\tau_j$
is diagonal in $r$, so the joint output is classical on $J$.
Its $BC$ state is the fixed pure state $|0\rangle\langle0|_{BC}$
in product with $\Omega_{AJD}$. Hence $I(A;BC\mid JD)_\Omega=0$,
which proves \ref{assert:JD-exists}).
\end{proof}

\section{Application to double Markov condition}
\label{sec:double-markov-applications}

\subsection{Classical conditional double Markovity}
\label{sec:diagonal-conditioning}
As discussed in \cite{HayashiZhao2026DoubleMarkov}, the classical double Markovity considers a joint distribution $p_{ABCD}$ of four classical random variables $A,B,C,D$ satisfying the double Markov conditions. We now consider a slightly more general setting, where $A$ may be quantum, while $B,C,D$ are classical:
\begin{equation}
 \rho_{ABCD}=\sum_{b,c,d}p_{bcd}\rho_{A|bcd}
             \otimes|bcd\rangle\langle bcd|.
 \label{eq:diagonal-joint-state}
\end{equation}
\begin{proposition}
\label{prop:diagonal-double-markov}
For \eqref{eq:diagonal-joint-state}, the double Markov conditions
\eqref{eq:quantum-crossed-conditions} hold if and only if there are a
finite classical variable $J$ and functions $g_{BD},g_{CD}$ such that
\begin{equation}
 \begin{gathered}
 J=g_{BD}(B,D)=g_{CD}(C,D)\quad\text{almost surely},
 \end{gathered}
 \label{eq:diagonal-common-label}
\end{equation}
and $ I(A;BC\mid DJ)=0$.
\end{proposition}

\begin{proof}
Assume \eqref{eq:quantum-crossed-conditions}. By
Lemma~\ref{lem:common-support-regularization}, we may compute
$\mathcal M_{\mathrm{com}}$ using the regularized conditional states
$\widetilde\rho_{Y\mid F}$. The joint state in
\eqref{eq:diagonal-joint-state} is diagonal on $Y=BCD$, so both
$\widetilde\rho_{Y\mid F}$ and $\rho_Y$ are diagonal in the basis
$\{|bcd\rangle:p_{bcd}>0\}$ of $P_Y\mathcal H_Y$. Their relative
cocycles are therefore diagonal. By Lemma~\ref{lem:generated-algebra},
$\mathcal M_{\mathrm{com}}$ is a commutative algebra of diagonal operators.

Let $\{\Pi_j\}_j$ be its minimal central projections. Then
\begin{equation}
 \mathcal M_{\mathrm{com}}=\operatorname{span}\{\Pi_j\}_j,
 \qquad \sum_j\Pi_j=P_Y.
 \label{eq:diagonal-common-algebra}
\end{equation}
For $y=(b,c,d)$ with $p_{bcd}>0$, let $j(y)$ be the unique index
satisfying $\Pi_{j(y)}|y\rangle=|y\rangle$. Define $J=j(B,C,D)$.

To show that either observation determines $J$, fix $R\in\{BD,CD\}$.
By Theorem~\ref{thm:main-common-information}, the preimages
\[
 \Pi_{j,R}:=
 \bigl(\mathcal T_R^*|_{\mathcal M(\mathsf S_R)}\bigr)^{-1}(\Pi_j)
\]
are orthogonal projections summing to $P_R$. They are diagonal because
the observed conditional states and their reference state are diagonal,
so their cocycles generate a diagonal algebra. For an occupied $y$,
write $r$ for the corresponding outcome $(b,d)$ or $(c,d)$. Then
\begin{align*}
 \langle r|\Pi_{j,R}|r\rangle
 &=\langle y|\mathcal T_R^*(\Pi_{j,R})|y\rangle\\
 &=\langle y|\Pi_j|y\rangle=\delta_{j,j(y)}.
\end{align*}
Thus the unique projection $\Pi_{j,R}$ containing $|r\rangle$
determines $j(y)$ from $r$ alone. This gives both functions in
\eqref{eq:diagonal-common-label} on the occupied outcomes.

To derive $I(A;BC\mid DJ)=0$, record the label in a register with
orthonormal basis $\{|j\rangle_J\}_j$. In tensor order $Y,J$, define
\begin{equation}
 U\xi=\sum_j\Pi_j\xi\otimes|j\rangle_J,
 \qquad \xi\in P_Y\mathcal H_Y.
 \label{eq:diagonal-recording-isometry}
\end{equation}
Since the $\Pi_j$ are orthogonal and sum to $P_Y$, $U$ is an isometry
on the input support. In tensor order $BC,J,D$, it sends $|bcd\rangle$ to
$|bc\rangle\otimes|j(b,c,d)\rangle_J\otimes|d\rangle$.
The output
$\Omega=(I_A\otimes U)\rho_{AY}(I_A\otimes U^*)$ is classical on $J$
and records $J$ without changing the marginal on $ABCD$.

Put $P_{JD}=\operatorname{supp}\Omega_{JD}$. On the generators
$\Pi_j$ of $\mathcal M_{\mathrm{com}}$, set
\[
 \pi_U(\Pi_j)=P_{JD}(|j\rangle\langle j|_J\otimes I_D)P_{JD}.
\]
The marginal $\Omega_{JD}$ is block diagonal on $J$, so its
support projection $P_{JD}$ commutes with each
$|j\rangle\langle j|_J\otimes I_D$. Thus the compressed operators
are orthogonal projections summing to $P_{JD}$. Each is nonzero
because $\Pi_j$ contains an occupied basis vector. Their
linear extension is therefore an injective $*$-representation into
$P_{JD}(\mathcal D_J\otimes\mathcal L(\mathcal H_D))P_{JD}$.
For every occupied $y$,
\[
 (I_{BC}\otimes\pi_U(\Pi_j))U|y\rangle
 =\delta_{j,j(y)}U|y\rangle
 =U\Pi_j|y\rangle.
\]
Thus \eqref{eq:JD-compatible-realization} holds on the input support.
Theorem~\ref{thm:classical-quantum-realization} now gives
$I(A;BC\mid DJ)_\Omega=0$.

Conversely, suppose that $J$ satisfies
\eqref{eq:diagonal-common-label} and $I(A;BC\mid DJ)=0$.
Since $J$ is a function of $(B,D)$, adjoining it to the conditioning
system $BD$ does not change $I(A;C\mid BD)$. By the chain rule
and nonnegativity, $I(A;C\mid BDJ)\leq I(A;BC\mid DJ)=0$.
Using that $J$ is also a function of $(C,D)$ gives
$I(A;B\mid CD)=0$ in the same way.
\end{proof}

\begin{remark}
For occupied outcomes $y$ and $y'$, the label constructed in the
proof of Proposition~\ref{prop:diagonal-double-markov} satisfies
$j(y)=j(y')$ if and only if $\rho_{A\mid y}=\rho_{A\mid y'}$.
Thus the minimal central projections give the equivalence classes of
conditional states used in the classical proof
\cite[Lemma~1]{HayashiZhao2026DoubleMarkov}.

Indeed, writing $p_y=p_{bcd}$ for $y=(b,c,d)$, the diagonal entry of
the regularized state is
\[
 \langle y|\widetilde\rho_{Y\mid F}|y\rangle
 =\frac{p_y}{2}\left(1+\frac{\Tr(\rho_{A\mid y}F)}{p_F}\right)
 \qquad(F\in\mathcal F_+).
\]
The diagonal entry of
$\widetilde\rho_{Y\mid F}^{it}\rho_Y^{-it}$ at $y$ is the
$it$-power of the ratio of this entry to $p_y$. Now $j(y)=j(y')$
holds exactly when the diagonal entries at $y$ and $y'$ agree
for every element of $\mathcal M_{\mathrm{com}}$, or equivalently for
all its cocycle generators. Equality of their entries for every
$t\in\mathbb R$ is equivalent to equality of the underlying positive
ratios. Hence equality of the labels is equivalent to
$\Tr(\rho_{A\mid y}F)=\Tr(\rho_{A\mid y'}F)$ for every
$F\in\mathcal F_+$. These effects separate the conditional states,
which proves the characterization of $j$.
\end{remark}

\begin{remark}\label{rmk:U}
For the label constructed in the proof of
Proposition~\ref{prop:diagonal-double-markov}, one also has
$I(A;BCD\mid J)_\Omega=0$. The preceding remark shows that
$\rho_{A\mid y}$ is the same for every occupied $y=(b,c,d)$ with
$j(y)=j$. Hence, in the conditional state given $J=j$, this common
state on $A$ factors out of the mixture over $y$. The conditional
state on $A$ and $BCD$ is a product, which gives the claimed equality.
\end{remark}

\subsection{A counterexample}
\label{sec:four-cycle-main}

Proposition~\ref{prop:diagonal-double-markov} records a label $J$
while preserving the state on $ABCD$. The following example
shows that an extension with this property can fail to give
$I(A;BC\mid DJ)=0$ when $B,C,D$ are quantum.

Let $A,B,C,D$ be qubits. In tensor order $A,B,D,C$, define
\begin{align}
 |G\rangle&=\frac14\sum_{a,b,d,c\in\{0,1\}}
 (-1)^{ab+bd+dc+ca}|a,b,d,c\rangle,
 \\ \rho_{ABCD}&=|G\rangle\langle G|.
 \label{eq:main-cycle-explicit-state}
\end{align}

\begin{proposition}[Conditional mutual informations of the four-qubit state]
\label{prop:four-cycle-CMI}
The state in \eqref{eq:main-cycle-explicit-state} satisfies
\begin{align}
 I(A;C\mid BD)_\rho&=0,
 &I(A;B\mid CD)_\rho&=0,\notag\\
 I(A;BC\mid D)_\rho&=1.
 \label{eq:main-cycle-CMI}
\end{align}
For every state $\Omega_{ABCDJ}$ with
$\Tr_J\Omega_{ABCDJ}=\rho_{ABCD}$, one has
$I(A;BC\mid DJ)_\Omega=1$.
\end{proposition}

\begin{proof}
Put the Pauli matrix
\[
X=\begin{pmatrix}0&1\\1&0\end{pmatrix},\qquad
Y=\begin{pmatrix}0&-i\\i&0\end{pmatrix},\qquad
Z=\begin{pmatrix}1&0\\0&-1\end{pmatrix}.
\]
Taking partial traces in the displayed form of $|G\rangle$ gives,
in the respective computational bases,
\begin{align}
  \rho_A = \rho_B = \rho_C = \rho_D &= \frac{I_2}{2}, \notag\\
 \rho_{AB}=\rho_{AC}=\rho_{BD}=\rho_{CD}&=\frac{I_4}{4},\notag\\
 \rho_{AD}=\rho_{BC}&=\frac{I_4+X\otimes X}{4}.
 \label{eq:main-cycle-explicit-marginals}
\end{align}

Both $\rho_{AD}$ and $\rho_{BC}$ in
\eqref{eq:main-cycle-explicit-marginals} have eigenvalues
$1/2,1/2,0,0$. Hence
\begin{align}
 S(A)=S(B)=S(C)=S(D)&=1,\notag\\
 S(AB)=S(AC)=S(BD)=S(CD)&=2,\notag\\
 S(AD)=S(BC)&=1.
 \label{eq:main-cycle-basic-entropies}
\end{align} 
For the marginal of three systems, grouping the terms of \(|G\rangle\) by the value of \(D\) gives
\begin{align*}
|G\rangle&=\frac{|\psi_0\rangle_{ABC}|0\rangle_D+
                     |\psi_1\rangle_{ABC}|1\rangle_D}{\sqrt2},
\\
\text{where }|\psi_d\rangle_{ABC}
&=\frac{|d,+,+\rangle+|(1-d),-,-\rangle}{\sqrt2}.
\end{align*}

Tracing out \(D\) removes the cross terms, so
\[
\rho_{ABC}
=\tfrac12|\psi_0\rangle\langle\psi_0|
+\tfrac12|\psi_1\rangle\langle\psi_1|.
\]
Since the joint state is pure, each three-qubit marginal has the
same nonzero eigenvalues as its one-qubit complement. All four
one-qubit marginals are maximally mixed, so
\[S(ABC) = S(ABD) = S(ACD) = S(BCD) = 1.\]
Hence,
\begin{align*}
 I(A;C\mid BD)_\rho=I(A;B\mid CD)_\rho&=1+1-2-0=0,\\
 I(A;BC\mid D)_\rho&=1+1-1-0=1.
\end{align*}

Now let $\Omega_{ABCDJ}$ have marginal $\rho_{ABCD}$ on $ABCD$.
The support of $\Omega_{ABCDJ}$ is contained in
$\operatorname{span}\{|G\rangle\}\otimes\mathcal H_J$, since its
marginal on $ABCD$ is pure. Thus
$\Omega_{ABCDJ}=\rho_{ABCD}\otimes\Omega_J$, and conditioning on
the independent register $J$ gives
$I(A;BC\mid DJ)_\Omega=I(A;BC\mid D)_\rho=1$.
\end{proof}

We next identify the common sufficient algebra and compare it with
operators on the original subsystem $D$. We will show that
\begin{align}\label{counter-algebra}
  \mathcal M_{\mathrm{com}} = M_2.
\end{align}
Since $\rho_A=I_2/2$, we can write
$|G\rangle=2^{-1/2}\sum_{a=0}^1 |a\rangle_A|u_a\rangle_Y$.
The two vectors
$|u_a\rangle_Y=\sqrt{2}(\langle a|_A\otimes I_Y)|G\rangle$
form an orthonormal basis of $P_Y\mathcal H_Y$.
In this basis on the support of $Y$, the joint state is
\[
\rho_{AY}=\frac12
\begin{pmatrix}
1&0&0&1\\
0&0&0&0\\
0&0&0&0\\
1&0&0&1
\end{pmatrix}.
\]
From this matrix, \(\rho_Y=I_2/2\), and for a nonzero effect \(F\) on \(A\),
\[
\rho_{Y\mid F}
=\frac{\operatorname{Tr}_A[(F\otimes I_2)\rho_{AY}]}{\operatorname{Tr}F/2}
=\frac{F^{\mathsf T}}{\operatorname{Tr}F}.
\]
For each Pauli matrix $\sigma\in\{X,Y,Z\}$, choose the effect
$F=(I_2+\sigma^{\mathsf T}/2)/2$. It has trace one and gives
$\rho_{Y\mid F}=(I_2+\sigma/2)/2$. The cocycle at $t=1$ is
$(I_2+\sigma/2)^i$. The ratio of its two eigenvalues is $3^i\ne 1$,
so it is a linear combination of $I_2$ and $\sigma$
with nonzero coefficient of $\sigma$. The algebra in
Lemma~\ref{lem:generated-algebra} contains the identity and this
cocycle, so it contains $\sigma$. Thus it contains the Pauli basis
$\{I_2,X,Y,Z\}$ of $M_2$, which gives \eqref{counter-algebra}.

In the same basis, an operator $x$ on the original $D$
has compressed matrix entries
\[
 \langle u_a|(I_{BC}\otimes x)|u_b\rangle
 =2\langle b|\Tr_D[(I_A\otimes x)\rho_{AD}]|a\rangle .
\]
By \eqref{eq:main-cycle-explicit-marginals}, the compressed matrix is
\[
 \tfrac12\Tr(x) I_2+\tfrac12\Tr(xX)X .
\]
The compressed $D$-local operators thus form the two-dimensional
space spanned by $I_2$ and $X$. In particular, they cannot
represent all of $\mathcal M_{\mathrm{com}}=M_2$ on the original
support.

The representation theorem also allows the output factors to be
reassigned by an isometry. To see the effect here, use output
systems $B',C',D'$ of the same dimensions as $B,C,D$, take $J$
one dimensional, and define the isometry on the occupied input by
\[
 U|u_a\rangle=|00\rangle_{B'C'}\otimes|0\rangle_J
                  \otimes |a\rangle_{D'} \qquad (a=0,1).
\]
The output has $B'C'$ in a fixed pure state in product with $AD'J$,
so $I(A;B'C'\mid D'J)_\Omega=0$. Here the output $D'$ holds the
two-dimensional quantum state from the entire input support of
$BCD$. In particular, the output marginal on $AB'C'D'$ differs
from $\rho_{ABCD}$.

\subsection{Tripartite quantum double Markovity}
\label{sec:tripartite-DM}

In the following two subsections, we use Theorem \ref{thm:classical-quantum-realization} to recover two main theorems in \cite{HayashiZhao2026DoubleMarkov}.

\begin{proposition}[c.f. {\cite[Theorem~1]{HayashiZhao2026DoubleMarkov}}]
\label{thm:tripartite-double-Markov}
For a state $\rho_{ABC}$, the conditions
\begin{equation}
 I(A;C\mid B)_\rho=I(A;B\mid C)_\rho=0
 \label{eq:tripartite-crossed-conditions}
\end{equation}
hold if and only if there are projective measurements on $B$ and $C$
whose separate recordings produce the same state $\Omega_{ABCJ}$,
classical on $J$, with
\begin{equation}
 \Tr_J\Omega=\rho_{ABC},\qquad I(A;BC\mid J)_\Omega=0.
 \label{eq:TH1-common-label-properties}
\end{equation}
\end{proposition}

\begin{proof}
Assume \eqref{eq:tripartite-crossed-conditions}. By
Theorem~\ref{thm:main-common-information} with $D$ one dimensional,
for each $R=B,C$ the restriction of $\mathcal T_R^*$ to
$\mathcal M(\mathsf S_R)$ is a $*$-isomorphism onto
$\mathcal M_{\mathrm{com}}$. Let $Z$ be the other system in $Y=BC$.
For $x\in\mathcal M(\mathsf S_R)$, extend $x$ by zero outside
$P_R\mathcal H_R$. Multiplicativity gives
\begin{align*}
 &P_Y(x^*\otimes I_Z)(I_Y-P_Y)(x\otimes I_Z)P_Y\\
 &\quad=\mathcal T_R^*(x^*x)
 -\mathcal T_R^*(x)^*\mathcal T_R^*(x)=0.
\end{align*}
The left-hand side is $K^*K$ with
$K=(I_Y-P_Y)(x\otimes I_Z)P_Y$, so $K=0$. Thus $x\otimes I_Z$
preserves $P_Y\mathcal H_Y$ and acts there as $\mathcal T_R^*(x)$.
The operators $b\otimes I_C$ and $I_B\otimes c$ commute on the
full space. Thus, for $b\in\mathcal M(\mathsf S_B)$ and
$c\in\mathcal M(\mathsf S_C)$,
\begin{equation}
 [\mathcal T_B^*(b),\mathcal T_C^*(c)]
 =P_Y[b\otimes I_C,I_B\otimes c]P_Y=0.
 \label{eq:quantum-compressed-commutation}
\end{equation}
Both restricted maps are onto $\mathcal M_{\mathrm{com}}$, so
$\mathcal M_{\mathrm{com}}$ is commutative.

Let $\{\Pi_j\}_j$ be the minimal central projections of
$\mathcal M_{\mathrm{com}}$. Then
\begin{equation}
 \mathcal M_{\mathrm{com}}=\operatorname{span}\{\Pi_j\}_j,
 \qquad \sum_j\Pi_j=P_Y .
 \label{eq:tripartite-common-projections}
\end{equation}
By the bimodule property of $\mathbb E_Y$, for $x\in\mathcal N_Y$,
\[
\begin{aligned}
 \mathbb E_Y\left(\sum_j\Pi_j x\Pi_j\right)
 &=\sum_j\Pi_j\mathbb E_Y(x)\Pi_j\\
 &=\mathbb E_Y(x).
\end{aligned}
\]
The last equality uses that $\mathbb E_Y(x)$ belongs to
$\mathcal M_{\mathrm{com}}$ and the $\Pi_j$ sum to $P_Y$.
The map $x\mapsto\sum_j\Pi_j x\Pi_j$ is its own trace adjoint.
Thus taking adjoints shows that this map fixes every output of
$\mathbb E_Y^*$. Applying it on $Y=BC$ in
\eqref{eq:common-algebra-joint-state-preservation} gives
\begin{equation}
 \rho_{ABC}
 =\sum_j(I_A\otimes\Pi_j)\rho_{ABC}(I_A\otimes\Pi_j).
 \label{eq:tripartite-input-blocks}
\end{equation}

Let $J$ have an orthonormal basis $\{|j\rangle_J\}_j$ indexed by
these projections. As in \eqref{eq:diagonal-recording-isometry},
define
\begin{equation}
 \begin{aligned}
 U\xi&=\sum_j\Pi_j\xi\otimes|j\rangle_J,
       \qquad \xi\in P_Y\mathcal H_Y,\\
 \pi_U(\Pi_j)&=|j\rangle\langle j|_J .
 \end{aligned}
 \label{eq:tripartite-isometry-representation}
\end{equation}
By \eqref{eq:tripartite-input-blocks}, the output is
\[
 \Omega_{ABCJ}
 =\sum_j (I_A\otimes\Pi_j)\rho_{ABC}(I_A\otimes\Pi_j)
       \otimes |j\rangle\langle j|_J .
\]
Thus $\Omega$ is classical on $J$ and $\Tr_J\Omega=\rho_{ABC}$.
Each label has probability $\Tr(\rho_Y\Pi_j)>0$, so $\pi_U$ is
an injective $*$-representation with unit $P_{JD}=I_J$ under
the identification of $JD$ with $J$. Since
$(I_{BC}\otimes\pi_U(\Pi_j))U=U\Pi_j$ for every $j$,
\eqref{eq:JD-compatible-realization} holds. By
Theorem~\ref{thm:classical-quantum-realization},
$I(A;BC\mid J)_\Omega=0$.

It remains to show that the same $U$ can be obtained from a
projective measurement on either $B$ or $C$. For $R=B,C$, let
$\Pi_{j,R}$ be the preimage of $\Pi_j$ under the restricted
$\mathcal T_R^*$. These are orthogonal projections summing to
$P_R$. By the support calculation above,
\[
 (\Pi_{j,R}\otimes I_Z)\xi=\Pi_j\xi
 \qquad(\xi\in P_Y\mathcal H_Y).
\]
Summing over $j$ gives $(P_R\otimes I_Z)\xi=\xi$. We may therefore
complete the projections to a projective measurement on
$\mathcal H_R$ by adding $I_R-P_R$ to one of them. This addition
vanishes on the occupied input. Hence the recording on either
$B$ or $C$ acts as $U$ there and produces the same $\Omega$.

Conversely, suppose the two projective recordings give an $\Omega$
satisfying \eqref{eq:TH1-common-label-properties}. For either
measurement, let $Q_j$ denote its projections lifted to $ABC$ by
tensoring with identities. The condition $\Tr_J\Omega=\rho_{ABC}$
then gives $\sum_j Q_j\rho_{ABC}Q_j=\rho_{ABC}$. It follows that
$Q_j\rho_{ABC}Q_k=0$ for $j\ne k$. Thus the isometry that appends
$|j\rangle_J$ on the range of $Q_j$ also produces $\Omega$, since
the terms between different labels vanish. For the measurement on
$B$, this is a local isometry from $B$ to $BJ$, so
$I(A;C\mid B)_\rho=I(A;C\mid BJ)_\Omega$. For the measurement on $C$,
it gives $I(A;B\mid C)_\rho=I(A;B\mid CJ)_\Omega$. Each conditional
mutual information on the right vanishes by the chain rule and
nonnegativity, since $I(A;BC\mid J)_\Omega=0$.
\end{proof}
\begin{remark}
The projections on the occupied spaces of $B$ and $C$ are
determined up to relabeling by the minimal central projections
of $\mathcal M_{\mathrm{com}}$ and the two isomorphisms in
Theorem~\ref{thm:main-common-information}.
\end{remark}

\subsection{Faithful-marginal double Markovity}
\label{sec:full-support-intersection}

The following theorem originally proved in \cite{HayashiZhao2026DoubleMarkov} used a careful block analysis of decompositions of joint state. However, from the perspective of sufficient algebra, the proof is rather short.

\begin{proposition}[c.f. {\cite[Theorem~2]{HayashiZhao2026DoubleMarkov}}]
\label{thm:full-support-conditional-intersection}
Suppose $\rho_{BCD}$ is full rank on $\mathcal H_{BCD}$. Then
\begin{equation}
 \begin{gathered}
 I(A;C\mid BD)_\rho=I(A;B\mid CD)_\rho=0\\
 \Longleftrightarrow\quad I(A;BC\mid D)_\rho=0 .
 \end{gathered}
 \label{eq:extended-intersection-conclusion-revised}
\end{equation}
\end{proposition}

\begin{proof}
The sufficiency follows from the chain rule and nonnegativity.
We show the necessity.

Assume \eqref{eq:quantum-crossed-conditions}. Since $\rho_Y$
is full rank on $Y=BCD$, its $BD$ and $CD$ marginals are also full
rank. Thus $P_Y=I_Y$, and each supported adjoint $\mathcal T_R^*$
acts by tensoring with an identity on the omitted system. In tensor
order $B,D,C$, Theorem~\ref{thm:main-common-information} gives
\begin{align}
 \mathcal M_{\mathrm{com}}
 &\subseteq
 \bigl(\mathcal L(\mathcal H_{BD})\otimes I_C\bigr)
 \cap \bigl(I_B\otimes\mathcal L(\mathcal H_{DC})\bigr)
 \notag\\
 &=I_B\otimes\mathcal L(\mathcal H_D)\otimes I_C.
 \label{eq:faithful-common-algebra-on-D}
\end{align}
The intersection consists of operators commuting with all
operators on $B$ and on $C$, which gives the last equality.

Take $J$ to be one dimensional with unit vector $|0\rangle_J$.
In the output order $BC,J,D$, define the isometry by
\[
 U|b,d,c\rangle=|bc\rangle\otimes|0\rangle_J\otimes|d\rangle.
\]
It maps one orthonormal basis onto the other and is therefore unitary
between the stated input and output spaces. The output $\Omega$ is
$\rho_{ABCD}$ with the one-dimensional register $J$ appended, so it is
classical on $J$. Since $\rho_D$ is full rank, $P_{JD}=I_J\otimes I_D$.

For $m=I_B\otimes x\otimes I_C\in\mathcal M_{\mathrm{com}}$,
put $\pi_U(m)=|0\rangle\langle0|_J\otimes x$.
This is an injective $*$-representation with unit $P_{JD}$.
The definition of $U$ gives
\[
 U^*(I_{BC}\otimes\pi_U(m))U=m.
\]
Since $UU^*=I_{BCJD}$, multiplying this identity by $U$ on the left
gives $(I_{BC}\otimes\pi_U(m))U=Um$. Theorem~
\ref{thm:classical-quantum-realization} now yields
$I(A;BC\mid JD)_\Omega=0$. The one-dimensional register contributes
nothing to the conditioning, so $I(A;BC\mid D)_\rho=0$.
\end{proof}

\section{Discussion}\label{sec:discussion-conclusion}
The measurement-induced state family separates the information needed
to recover correlations with $A$ from other information in an
observation. Sufficient observations identify the same minimal algebra.
Its block structure gives a sufficient subsystem, and the
classical--quantum realization criterion determines which quantum blocks
fit in a prescribed output system.

If $\rho_{ABCD}$ satisfies the double Markov conditions and is
diagonal in product bases on $B,C,D$, the conditional density
operators of $A$ partition the triples $(b,c,d)$ of positive
probability into equality classes. Either $BD$ or $CD$ determines
the same class label $J$. Recording $J$ preserves the original
$ABCD$ marginal and gives $I(A;BC\mid DJ)_\Omega=0$. The conditional
states of $A$ may be noncommuting.

The four-qubit example has a noncommutative common algebra,
$\mathcal M_{\mathrm{com}}=M_2$. Every extension $\Omega_{ABCDJ}$
with the original $ABCD$ marginal has $I(A;BC\mid DJ)_\Omega=1$.
The observables on physical $D$ span a two-dimensional space
after compression to the support of $\rho_{BCD}$. The isometry
in Section~\ref{sec:four-cycle-main} realizes the algebra on the
output qubit $D'$. Its output has $B'C'$ in the fixed pure state
$|00\rangle_{B'C'}$ and $J$ is one-dimensional, giving
$I(A;B'C'\mid D'J)_\Omega=0$.

Under the double Markov conditions, if $D$ is one-dimensional,
separate projective measurements on $B$ and on $C$ give the same
recorded state $\Omega_{ABCJ}$ with the original $ABC$ marginal
and $I(A;BC\mid J)_\Omega=0$. With the double Markov conditions and
a full-rank $\rho_{BCD}$, the common algebra acts on the physical
overlap $D$ and $I(A;BC\mid D)_\rho=0$.

The results concern exact preservation of a specified state.
Approximate recovery from small conditional mutual information is
known \cite{FawziRenner2015,SutterFawziRenner2016}. An approximate
counterpart would also have to control changes in the sufficient
algebra and its realizations on the specified regions.

\section*{Acknowledgment}
MH and JZ would like to thank Prof. Chandra Nair
for fruitful discussions.
JZ would also like to express his gratitude to
Prof. Chandra Nair, his supervisor, for introducing
the problem framework and motivation, sharing
proofs from classical information theory, and
providing valuable insights throughout the research.

An initial draft of this manuscript was prepared
using AI (Microsoft Copilot Enterprise and Chat GPT Pro)
based on the research questions and ideas
of the authors. The authors subsequently rewrote
substantial parts of the manuscript, revising the
mathematical formulation and restructuring the
logical development of the arguments.
Throughout this process, the authors carefully
verified the mathematical content and proofs.
The authors take full responsibility for the
final manuscript.

\bibliographystyle{IEEEtran}
\bibliography{mybiblio}

@article{Petz1986SufficientSubalgebras,
  author  = {Petz, D{\'e}nes},
  title   = {Sufficient Subalgebras and the Relative Entropy of States of a von Neumann Algebra},
  journal = {Communications in Mathematical Physics},
  volume  = {105},
  number  = {1},
  pages   = {123--131},
  year    = {1986},
  doi     = {10.1007/BF01212345},
}

@article{HJP,
   title={Structure of States Which Satisfy Strong Subadditivity of Quantum Entropy with Equality},
   volume={246},
   ISSN={1432-0916},
   doi={10.1007/s00220-004-1049-z},
   number={2},
   journal={Communications in Mathematical Physics},
   publisher={Springer Science and Business Media LLC},
   author={Hayden, Patrick and Jozsa, Richard and Petz, Denes and Winter, Andreas},
   year={2004},
   month=apr, pages={359--374} }

@article{JencovaPetz2006Sufficiency,
 author={Jen{\v c}ov{\'a}, Anna and Petz, D{\'e}nes},
 title={Sufficiency in Quantum Statistical Inference},
 journal={Communications in Mathematical Physics}, volume={263}, number={1},
 pages={259--276}, year={2006}, doi={10.1007/s00220-005-1510-7}}

@misc{HayashiZhao2026DoubleMarkov,
  author = {Hayashi, Masahito and Zhao, Jinpei},
  title = {Double Markovity for quantum systems},
  year = {2026},
  eprint = {2601.09995},
  archivePrefix = {arXiv},
  primaryClass = {quant-ph},
  url = {https://arxiv.org/abs/2601.09995}
}

@article{FawziRenner2015,
  author = {Omar Fawzi and Renato Renner},
  title = {Quantum conditional mutual information and approximate Markov chains},
  journal = {Communications in Mathematical Physics},
  volume = {340},
  pages = {575--611},
  year = {2015},
  doi = {10.1007/s00220-015-2466-x}
}

@article{SutterFawziRenner2016,
  author = {David Sutter and Omar Fawzi and Renato Renner},
  title = {Universal recovery map for approximate Markov chains},
  journal = {Proceedings of the Royal Society A},
  volume = {472},
  pages = {20150623},
  year = {2016}
}

@article{GengNair14,
 author = {Geng, Y. and Nair, C.},

 journal = {IEEE Transactions on Information Theory},
 number = 4,
 pages = {2087--2104},
 publisher = {IEEE},
 title = {The capacity region of the two-receiver {G}aussian vector broadcast channel with private and common messages},
 volume = 60,
 year = 2014,
 doi = {10.1109/TIT.2014.2304457}}

@article{ajn19,
  author = {Anantharam, Venkat and Jog, Varun and Nair, Chandra},
  title = {Unifying the {Brascamp--Lieb} Inequality and the Entropy Power Inequality},
  journal = {IEEE Transactions on Information Theory},
  volume = {68},
  number = {12},
  pages = {7665--7684},
  year = {2022},
  doi = {10.1109/TIT.2022.3192913}
}

@article{Shirokov2016,
  author  = {M. E. Shirokov},
  title   = {Measures of correlations in infinite-dimensional quantum systems},
  journal = {Sbornik: Mathematics},
  volume  = {207},
  number  = {5},
  pages   = {724--768},
  year    = {2016},
  doi     = {10.1070/SM8561}
}

@article{Dawid1980ConditionalOperations,
  author = {Dawid, A. Philip},
  title = {Conditional Independence for Statistical Operations},
  journal = {The Annals of Statistics},
  volume = {8},
  number = {3},
  pages = {598--617},
  year = {1980},
  doi = {10.1214/aos/1176345011}
}

@article{SanMartinMouchartRolin2005CommonInformation,
  author = {{San Mart\'in}, Ernesto and Mouchart, Michel and Rolin, Jean-Marie},
  title = {Ignorable Common Information, Null Sets and {Basu}'s First Theorem},
  journal = {Sankhy\=a: The Indian Journal of Statistics},
  volume = {67},
  number = {4},
  pages = {674--698},
  year = {2005}
}

@article{GacsKorner1973CommonInformation,
  author = {G\'acs, P\'eter and K\"orner, J\'anos},
  title = {Common Information Is Far Less Than Mutual Information},
  journal = {Problems of Control and Information Theory},
  volume = {2},
  number = {2},
  pages = {149--162},
  year = {1973}
}

@article{KoashiImoto2002,
  author = {Koashi, Masato and Imoto, Nobuyuki},
  title = {Operations That Do Not Disturb Partially Known Quantum States},
  journal = {Physical Review A},
  volume = {66},
  number = {2},
  pages = {022318},
  year = {2002},
  doi = {10.1103/PhysRevA.66.022318},
  eprint = {quant-ph/0101144},
  archivePrefix = {arXiv},
  url = {https://arxiv.org/abs/quant-ph/0101144}
}

\end{document}